\documentstyle[osa,aps,prl,epsfig]{revtex}
\begin{document}
\twocolumn[\hsize\textwidth\columnwidth\hsize\csname
@twocolumnfalse\endcsname
\title{
Synthesis of VO$_2$ Nanowire and Observation of the
Metal-Insulator Transition$^{\ast}$}
\author{$^{\dagger}$Sungyoul Choi, Bong-Jun Kim, Yong-Wook Lee, Sun Jin Yun, and Hyun-Tak Kim}
\address{IT Convergence and Components Research Lab., ETRI, Daejeon 305-350, Republic of
Korea}
\date{June 29, 2006}
\maketitle{}
\begin{abstract}
We have fabricated crystalline nanowires of VO$_2$ using a new
synthetic method. A nanowire synthesized at 650$^{\circ}$C shows
the semiconducting behavior and a nanowire at 670$^{\circ}$C
exhibits the first-order metal-insulator transition which is not
the one-dimensional property. The temperature coefficient of
resistance in the semiconducting nanowire is 7.06 $\%$/K at 300 K,
which is higher than that of commercial bolometer.\\ \\

\end{abstract}
]
 One dimensional (1-D) nanostructure materials exhibit unique
 physical properties that differ from their bulk properties.
 It is due to a characteristic of the 1-D structure such as
 nanotubes, nanorods, and nanowires [1-3]. It is well-known that
  an abrupt metal-insulator transition (MIT) and a hysteresis
  behavior do not occur in 1-D structure. These are an advantage
  for a device application. Therefore, synthetic efforts for 1-D
  materials have been continued by many researchers, although
  synthesis of 1-D structures is very difficult.

The transition oxide material, VO$_2$, undergoes the structural
phase transition (SPT) from the monoclinic to the rutile
tetragonal structures near 340 K. It was revealed that the
first-order MIT is controlled by hole doping of a low density and
is not caused by the SPT; this demonstrated the Mott transition
[4].  VO$_2$ has a lot of applications such as electro-optic
switch, infrared bolometer, and the Mott first-order field effect
transistor (FET), etc. New ideas for the first-order MIT
transistor were disclosed by Kim and Kang [5] and Chudnovski et
al. [6]. For the fabrication of nanometer-scale Mott FET devices,
the synthesis of single-crystalline VO$_2$ nanowires was reported
[7]. Metastable VO$_2$ nanowire arrays were synthesized via an
ethylene glycol reduction approach [8].

\begin{figure}
\vspace{-0.5cm}
\centerline{\epsfysize=7.0cm\epsfxsize=8.0cm\epsfbox{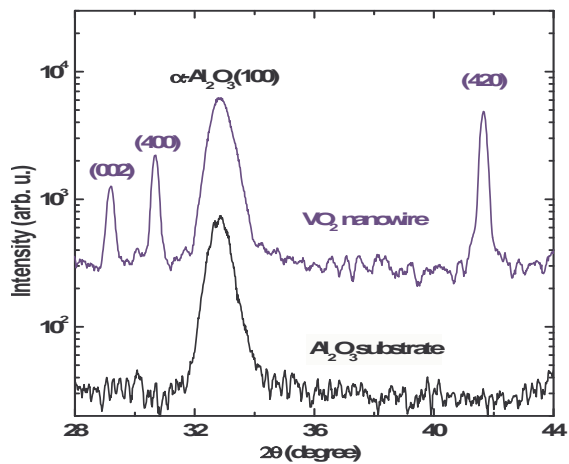}}
\vspace{-0.0cm} \caption{The crystal structure of a VO$_2$
nanowire grown at 670$^{\circ}$C for 30 minutes. XRD pattern of
the VO$_2$ nanowire compared to JCPDS.}
\end{figure}

In this paper, we reports synthesizing conditions of VO$_2$
nanowires fabricated by using a synthetic method. Their electrical
characteristics are analyzed by measuring the temperature
dependence of resistance and I-V characteristics. In particular,
to our knowledge, we first observed first-order MITs in nanowires.

\begin{figure}
\vspace{-1.0cm}
\centerline{\epsfysize=13.0cm\epsfxsize=9.0cm\epsfbox{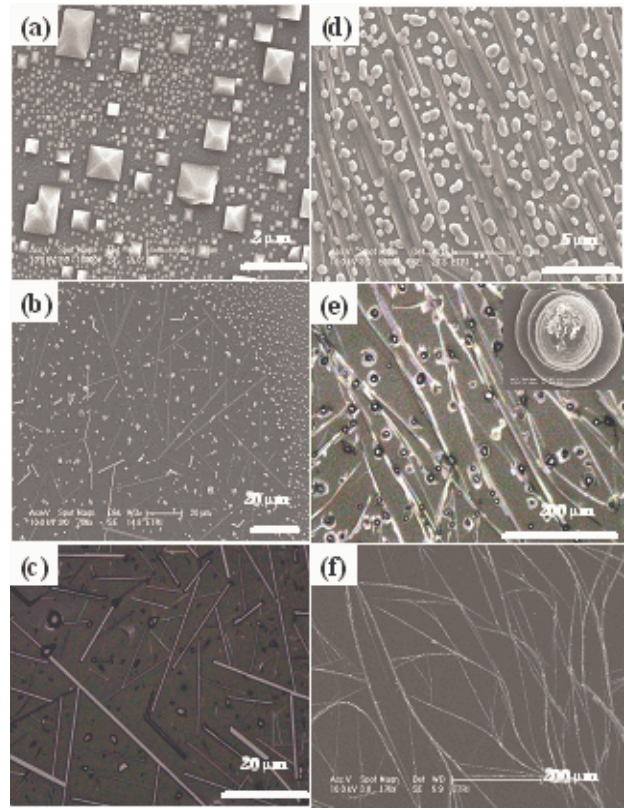}}
\vspace{-0.5cm} \caption{SEM images of VO$_2$ nanostructures. The
annealing temperatures and times are as follows: (a)
630$^{\circ}$C, (b) 650$^{\circ}$C, (c) 670$^{\circ}$C for 30 min,
and (d) 40 min, (e) 50 min, (f) 60 min at 670$^{\circ}$C.}
\end{figure}

VO$_2$ nanostructures with nanoblock and nanowire were
successfully synthesized by controlling the oxygen partial
pressure in the oxidation process of metallic vanadium. The
metallic vanadium was grown on $\alpha$-Al$_2$O$_3$ (01-10)
substrate at 500$^{\circ}$C in an Ar ambient atmosphere of 50
mTorr using RF Sputter. Advantage of this new method is a shorter
synthesis time than that of other nanowire fabrication methods
such as thermal chemical vapor deposition [9] and the bulk crystal
growth method [6]. The use of Al$_2$O$_3$ substrate different from
SiO$_2$/Si derived more high quality nanowire on the basis of the
fact that VO$_2$ film is well-grown on Al$_2$O$_3$. Annealing was
performed at 630$\sim$670$^{\circ}$C in the O$_2$ ambient
atmosphere of 50 mTorr for 30$\sim$60 minutes. Figure 1 shows an
X-ray diffraction (XRD) pattern of the crystal structure of a
VO$_2$ nanowire grown at 630$^{\circ}$C for 30 minutes. Lattice
constants from XRD peaks are calculated as a=12.03 $\AA$, b=6.693
$\AA$, c=6.42 $\AA$, which is in agreement with the reported
values of the monoclinic VO$_2$ for JCPDS (card No. 71-0042) [10].
(400) peak is the most intense peak of typical VO$_2$ thin film
grown on $\alpha$-Al$_2$O$_3$ substrate. The XRD peaks indicate
that the VO$_2$ nanowire is crystalline.

\begin{figure}
\vspace{-1.0cm}
\centerline{\epsfysize=13.0cm\epsfxsize=9.0cm\epsfbox{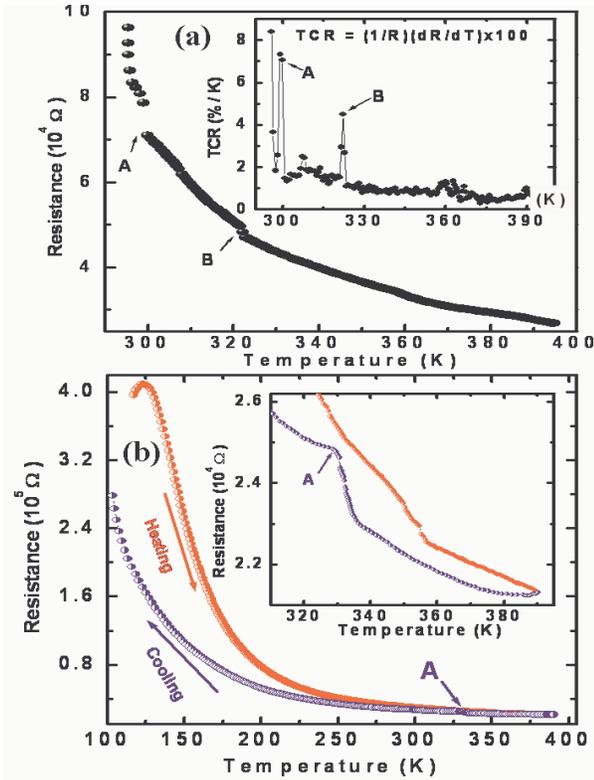}}
\vspace{-0.7cm} \caption{(a) Temperature dependence of resistance
of the VO$_2$ nanowire grown at 650$^{\circ}$C. Inset shows the
Temperature Coefficient of Resistance; TCR=-(1/R)(dR/dT)100. (b)
The hysteresis of temperature dependence of resistance of the
VO$_2$ nanowire. Inset is magnification of resistances around 330
K.}
\end{figure}

Figure 2 (a)$\sim$(c) show high resolution scanning electron
microscopy (SEM) images of nanostructures synthesized at several
annealing temperatures. The nanostructures are the semblance
nanoblocks (quadrangular pyramid) and nanowires. Nanoblocks with a
size of 50~500 nm were synthesized at 630$^{\circ}$C. Nanoblocks
and wires coexist at 650$^{\circ}$C. Only nanowires were grown at
670$^{\circ}$C. The nanowires in Fig. 2(c) are rectangular
parallelepiped form with a length of 10$\sim$800 $\mu$m (z-axis),
a width of 20$\sim$150 nm (y-axis) and a thickness of 100$\sim$500
nm (x-axis).

\begin{figure}
\vspace{-1.5cm}
\centerline{\epsfysize=13.0cm\epsfxsize=9.0cm\epsfbox{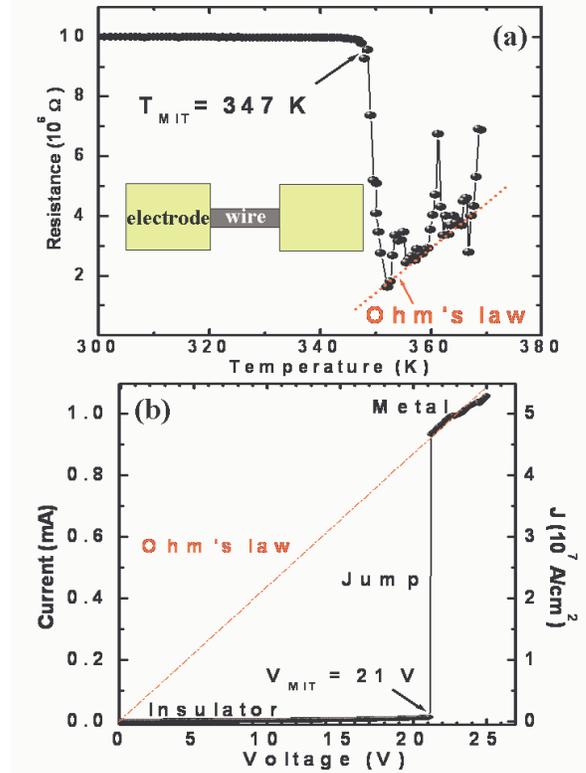}}
\vspace{-1.0cm} \caption{(a) Temperature dependence of resistance
of VO$_2$ nanowire of grown at 670$^{\circ}$C. Inset shows
electrode. (b) I-V measurement of VO$_2$ nanowire. It shows a
sharp current jump when an electrical field is applied to nanowire
with ~21 V.}
\end{figure}

Annealing time for fabrication of good nanowires was also changed
from 40 to 60 minutes with step of 10 minute, as shown in Fig. 2
(d)$\sim$(f). When annealing time was 40 min (Fig. 2(d)),
nanowires begin to connect with a neighbor nanowire, and it appear
a small nanoblock like a cone form. The inset in Fig. 2(e), it
shows a clearly cone form. Annealing time was 1 hour, a nanowire
is connected to other nanowires like a crooked bough, as shown in
Fig. 2(f). Thus, the optimum growth condition of VO$_2$ nanowires
on Al$_2$O$_3$ substrate is found to be at 670$^{\circ}$C and 30
min.

We measured the temperature dependence of resistance and I-V
characteristic curves to reveal electrical properties of VO$_2$
nanowires. Figure 3 shows the temperature dependence of resistance
and hysteresis curve for the VO$_2$ nanowires grown at
650$^{\circ}$C. The temperature dependence is semiconductive and
has no the MIT jump (abrupt change in resistance with temperature)
which is the resistance characteristic of VO$_2$. Moreover, the
temperature coefficient of resistance (TCR) has large values near
300 K and below 320 K (inset in Fig. 3 (a)). A TCR value at 300 K
is 7.06 $\%$/K, which is larger than that of a commercial
bolometer, as indicated by arrow A in Fig. 3(a). A TCR peak near
320 K is due to a change of the resistance near 320 K, as
indicated by arrow B. Figure 3(b) shows a lower resistance than
that at 300 K in Fig. 3(a) and the semiconducting temperature
behavior in the heating process. Hysteresis in the heating and
cooling process is exhibited. In the cooling process, a small step
of the resistance near 330 K appears, as indicated by arrow A in
the inset of Fig. 3(b), which can be regarded as the first-order
MIT in VO$_2$. This is attributed to hole excitation by heating
produced in the cooling process from a high temperature of 390 K,
on the grounds that the first-order MIT occurs by hole excitation
[4]. The fact that the magnitude of the jump is small is due to a
large resistance in the nanowire because the large resistance
reduces the magnitude of jump [5].

For the VO$_2$ nanowire grown at 670$^{\circ}$C, the sharp
first-order MIT jump near 347 K and the ohmic behavior above 347 K
are exhibited (Fig. 4(a)). The electric field-induced first-order
MIT is also measured (Fig. 4(b)). Jump of Current is
1.2$\times$10$^{-5}$ A to 9${\times}$10$^{-4}$ A at V$_{MIT}$=21V
and current follows Ohm's law in the larger voltage than
V$_{MIT}$. This indicates that the nanowire has a component of
metal. The MIT voltage can be controlled by varying the distance
between electrodes of nanowire. The observed first-order MITs are
attributed to breakdown of the critical on-site Coulomb
interaction between electrons [4,11].

In summary, we found conditions fabricating nanowires showing the
first-order MIT which is far from the 1-D structural property.
Furthermore, the crystalline nanowires with large resistnace are
useful to a lot of device applications.

\end{document}